\documentclass[aps, prd, showpacs, superscriptaddress, groupedaddress, twocolumn]{revtex4} 
\usepackage{graphicx}    
\usepackage{amssymb}
\usepackage{amsmath}
\usepackage{subfigure}
\usepackage[pagebackref=false, colorlinks=true]{hyperref}
\hypersetup{
linkcolor=blue,     
citecolor=blue,     
urlcolor=blue}   
\usepackage{soul}
\usepackage{amsmath}
\begin{document}
\title{Equilibrium states from gravitational collapse of minimally coupled scalar field with non-zero potential}
\author{Dipanjan Dey}
\email{deydipanjan7@gmail.com}
\affiliation{Department of Mathematics and Statistics,
Dalhousie University,
Halifax, Nova Scotia,
Canada B3H 3J5}
\author{Koushiki}
\email{koushiki.malda@gmail.com}
\affiliation{International Centre for Space and Cosmology, School of Arts and Sciences, Ahmedabad University, Ahmedabad, GUJ 380009, India}
\author{Pankaj S. Joshi}
\email{psjcosmos@gmail.com}
\affiliation{International Centre for Space and Cosmology, School of Arts and Sciences, Ahmedabad University, Ahmedabad, GUJ 380009, India}
\date{\today}

\begin{abstract}
We study the model of spherically symmetric and spatially homogeneous gravitational collapse of a minimally coupled scalar field. Our study focuses on obtaining the scalar field potential that leads to a final equilibrium state in the gravitational collapse. We demonstrate the existence of a class of scalar field solutions that can indeed result in such an end equilibrium state.\\
\\
$\textbf{key words}$: Gravitational collapse, Scalar field collapse, equilibrium condition.

\end{abstract}
\maketitle

\section{Introduction}
As predicted by the well-known singularity theorems \cite{Penrose}, the gravitational collapse of a massive
star, when it runs out of it's internal fuel, would result in a spacetime singularity where physical quantities 
diverge and the spacetime breaks down. Such singularities would be hidden within an event horizon
of gravity, forming a blackhole \cite{Oppenheimer_1939, Datt_1938}, or they could be cosmic singularities visible to faraway observers 
in the universe \cite{PSJ1, PSJ2, PSJ3, Mosani, Mosani2, Mosani3, Mosani4, Mosani5, DeyK1}. This is a major issue of the contemporary debate in black hole physics and 
cosmology today.

All the same, gravitational collapse is a fundamental process in nature. Cosmological structures 
such as galaxies, galactic clusters, and various structures in the universe form due gravitational 
collapse of matter clouds at various scales. In order to help the structure formation process,
the gravitational collapse of a massive matter cloud that was initiated at various epochs and scales 
in the universe must equilibrate or slow down considerably. There are different models of gravitational collapse where the stabilization of cosmic structures is explained. One of the frequently used models is the top-hat collapse model \cite{Gunn}. In the top-hat collapse model, the initial over-densities in our Universe are considered to be isolated from the external universe, and the dynamics of the over-dense regions are described by the closed Friedmann-Lemaître-Robertson-Walker (FLRW) spacetime which is in this model seeded by 
dustlike matter. In that model, the pressureless matter first expands with the background albeit at a slower rate, and at a certain time, due to its own gravitational pull, the expansion halts and the over-dense region starts collapsing. Now, since the matter is pressureless in the top-hat model, there is no way one can show any sign of stabilization of matter using the general relativistic Einstein equations. Therefore, in that model, the Newtonian virialization technique is used to stabilize the collapsing matter. Though in the top-hat collapse model, the Newtonian virialization technique is used in an ad-hoc way, this model produces theoretical results which closely resemble the observations. Since there exists no general relativistic definition of the Newtonian virialization process (at least as per our knowledge), this model is frequently used to describe cosmic structures. However, there are some gravitational collapse models where the collapsing matter asymptotically reaches an equilibrium state due to the effective non-zero pressure \cite{JMN11, Joshi:2013dva, Bhattacharya:2017chr, Dey:2019fja}. Recently, in \cite{Dey:2023qdt}, it is shown that the equilibrium condition can be obtained from the covariant counterpart of the virialization equation. However, the authors in that paper do not claim that the equilibrium condition is the general relativistic virialization condition. More research in this direction is necessary. It should be noted that in \cite{JMN11, Joshi:2013dva, Bhattacharya:2017chr, Dey:2019fja, Dey:2023qdt}, the entire dynamics of the gravitational collapse leading to the end equilibrium state are done within the framework of general relativity. Moreover, in \cite{Dey:2023qdt}, the authors demonstrate how this gravitational collapse model can be made cosmologically relevant. They also illustrate that this model can yield similar outcomes to those obtained from the top-hat collapse model.

In the present paper, we consider the matter constituent of the gravitational collapse model as a minimally coupled spatially homogeneous scalar field. We restrict ourselves to such fields because spatial inhomogeneity here would result in a type-II matter field, not observed physically \cite{hawking}. Our objective is to investigate the specific type of scalar field potential that could possibly give rise to the final equilibrium state. We demonstrate that a specific class of scalar field potentials enables the gravitational collapse of the scalar field to attain the end equilibrium state. In the type of collapsing dynamics described here, the non-zero negative pressure plays an important role. Since dark matter is generally considered to be pressureless, this type of equilibrium configuration can exist in a certain cosmological length scale where the effect of dark energy cannot be ignored. In \cite{ Maor:2005hq, Lahav:1991wc, Shapiro, Horellou:2005qc, Basilakos:2003bi, Basilakos:2006us, Basilakos:2009mz}, the authors extensively examine cosmological scenarios where the presence of dark energy influences the formation of dark matter structures. These studies reveal that considering the non-zero effect of dark energy can lead to variations in the virialization radius. In those studies, the authors utilized the Newtonian virialization technique. However, in the paper \cite{Dey:2023qdt}, the authors investigated similar scenarios by considering a gravitational collapse model leading to a final equilibrium state. In this paper, as mentioned above, we derive a class of scalar field potentials that exhibit similar collapsing dynamics as shown in \cite{Dey:2023qdt}. It is important to note that we do not propose the scalar field as dark matter or dark energy. Our focus is solely on obtaining the potential of the scalar field leading to an equilibrium, where the scalar field serves as the constituent of the collapsing matter. Therefore, the present work is primarily motivated by mathematical considerations, rather than cosmological ones. 

The structure of this paper is organized as follows. In Section (\ref{sec2}), we discuss the conditions necessary for the formation of a general relativistic equilibrium configuration, which arises as the final outcome of gravitational collapse. In Sec.~(\ref{sec3}), we obtain a class of minimally coupled scalar field potentials which can lead to an end equilibrium state of gravitational collapse. In Sec.~(\ref{sec4}), we take one example of such collapse dynamics which lead to an end equilibrium state and show the potential of the scalar field belongs to the class of potentials discussed in Sec.~(\ref{sec3}). We draw this example from \cite{Dey:2023qdt}. In Sec.~(\ref{sec5}), we conclude by discussing our results and their possible implications. Throughout the paper, we use a system of units in which the velocity of light and the universal gravitational constant (multiplied by $8\pi$), are both set equal to unity.

\section{End equilibrium state in gravitational collapse}
\label{sec2}
The metric of a general spherically symmetric collapsing system can be written as
\begin{eqnarray}
ds^2 = - e^{2\nu(r,t)} dt^2 + {R'^2\over G(r,t)}dr^2 + R^2(r,t) ~d\Omega^2\, ,
\label{genmetric}
\end{eqnarray}
where $\nu(r,t)$, $R(r,t)$ and $G(r,t)$ are functions of local coordinates $r$ and
$t$.  Since none of the functions depends upon angular coordinates, the metric given above can be used to describe a spherical
gravitational collapse. Due to the presence of the undetermined functions (i.e., $\nu(r,t)$, $R(r,t)$ and $G(r,t)$) of $r$ and $t$, the above metric can describe a large class of collapsing models of matter-fields. Here, a dot and a prime above any function are used to specify a derivative
of that function with respect to coordinate time and radius, respectively. As we know, the singularity theorems state that under a certain physically reasonable set of conditions such as retaining the causal regularity, non-violation of an energy condition on matter fields, and the existence of trapped surfaces, a spacetime singularity would be formed necessarily due to the catastrophic collapse of a general matter distribution. Due to the formation of trapped surfaces, the whole matter distribution ultimately sinks into that singular point. However, there exist various scenarios in gravitational collapse where trapped surfaces do not form due to the violation of strong energy conditions, or due to different evolution of geometries, and therefore the collapsing matter distribution may stop collapsing further, and ultimately reach an equilibrium state. 

In the realm of general relativity, a self-gravitating system can reach an equilibrium state \cite{JMN11, Joshi:2013dva, Bhattacharya:2017chr, Dey:2019fja}. The conditions for the equilibrium state of a spherical over-dense region of physical radius $R$ is
\begin{eqnarray}
\lim_{t\to \infty}\dot{R}=\lim_{t\to\infty}\ddot{R}=0\,\, ,
\label{eqlb}
\end{eqnarray}
which implies the collapsing system can reach the end equilibrium state in the asymptotic comoving time.
There exists a scaling degree of freedom in the metric mentioned above and using that freedom one can write \cite{JMN11}:
\begin{eqnarray}
R(r,t)= r~a(r,t)\,,
\label{scaling}
\end{eqnarray}
which implies at the equilibrium state:
\begin{eqnarray}
\dot{a}_e(r) =\ddot{a}_e(r) =0\,,
\label{stabc}
\end{eqnarray}
where we use subscript $e$ to denote the equilibrium value of any
quantity as
$$a_e(r) \equiv \lim_{t \to \infty} a(r,t)\, .$$
The above equilibrium condition can be redefined in a coordinate-invariant way using Cartan scalars. In an invariant frame, the frame derivatives $D\Psi_2$ and $\Delta \Psi_2$ are Cartan scalars, where the spin frame operators $D\equiv \ell^a \nabla_a\equiv o^A o^{A^{\prime}} \nabla_{A A^{\prime}}$, $\Delta \equiv n^a \nabla_a \equiv i^A i^{A^{\prime}} \nabla_{A A^{\prime}}$, where $(o^A, i^A)$ and $(o^{A^{\prime}}, i^{A^{\prime}})$ are the spin frame and its conjugate spin frame bases, respectively and $\Psi_2 =C_{abcd}l^a m^b\bar{m}^c n^d = C_{1342}$. Here, $C_{abcd}$ is the Weyl curvature tensor and $(l^a,  n^a, m^a, \bar{m}^a)$ are the corresponding null frame bases. The Cartan scalars $D\Psi_2$ and $\Delta \Psi_2$ can be combined to construct a new scalar. It can be shown that one can redefine the above equilibrium conditions by using a scalar $(D-\Delta)^2\Psi_2$ which can be constructed from $D\Psi_2$ and $\Delta \Psi_2$ \cite{Dey:2023qdt}. The redefined equilibrium condition is: 
\begin{eqnarray}
    \lim_{t \to \infty}(D-\Delta)^2\Psi_2 = 0~ \forall r.
\end{eqnarray}
The above definition of the equilibrium condition is coordinate-independent.

Since for a given comoving radius $r$, the scale factor $a(r,t)$ is a smooth monotonically decreasing function of comoving time $t$, the condition $\dot{a} = \ddot{a} = 0$ may be satisfied in a finite comoving time in some collapsing scenarios, however, that not as an equilibrium state. The condition $\dot{a} = \ddot{a} = 0$ at a finite comoving time ($t_e$), in general, does not imply an equilibrium state. An equilibrium state at the finite comoving time $t_e$ requires:
\begin{eqnarray}
   a(r,t) &>& a_e(r)~ \forall t\in [0,t_e) \implies \dot{a} < 0 ~ \forall t\in [0,t_e)\,\, ,\nonumber\\
 \text{and}~  a(r,t) &=& a_e(r)~ \forall t\geq t_e \implies \dot{a} = \ddot{a} = \dddot{a} =...\nonumber \\&=& a^{(n)} = 0 ~ \forall t\geq t_e \,\, . 
\end{eqnarray}
Therefore, if the collapsing system reaches the equilibrium state at a finite comoving time ($t_e$) then the first derivative of the scale factor $a(r, t)$ at that time becomes discontinuous (i.e., $\mathcal{C}^0$ function) which is not possible if we consider $g_{\mu\nu}(t,r)$ is at least $\mathcal{C}^2$. Therefore, for all possible scenarios of gravitational collapse, the equilibrium conditions can only be satisfied at the asymptotic comoving time.

\section{Homogeneous scalar field collapse}
\label{sec3}
Here, we consider a homogeneous scalar field $\phi (t)$ that seeds a closed FLRW spacetime. An FLRW spacetime is the appropriate choice as the metric given in Eq.(\ref{genmetric}) will become such in a homogeneous case.
The line element of FLRW spacetime is:
\begin{equation}\label{metric1}
    ds^2 = -dt^2 +\frac{a^2(t)}{1-k~r^2}dr^2+r^2 ~a^2(t)~ d\Omega^2.
\end{equation}
Here, $k = 1$ since we consider closed FLRW spacetime. 
Using the Einstein equations the energy density and pressure can be written as:
\begin{eqnarray}
\label{rho}
    \rho &=&\frac{\Dot{\phi}^2}{2}+V(\phi)=3\left(\frac{\Dot{a}^2}{a^2}+\frac{k}{a^2}\right)\,\, ,\\
    p &=&\frac{\Dot{\phi}^2}{2}-V(\phi)=-2\frac{\Ddot{a}}{a}-\frac{\Dot{a}^2}{a^2}-\frac{k}{a^2}\,\, ,
    \label{p}
\end{eqnarray}
where $V(\phi)$ is the potential of scalar field. At the equilibrium state, the energy density ($\rho_{eq}$) and pressure ($p_{eq}$) become,
\begin{eqnarray}
    \rho_{e}&=&\frac{3k}{a_{e}^2}\,\, ,\\
    p_{e} &=& -\frac{k}{a_{e}^2},
\end{eqnarray}
which implies that at the equilibrium state, the equation of state ($\omega_\phi$) of the scalar field should be: $\omega_\phi = -\frac13$. One can also verify that at the equilibrium state:
\begin{eqnarray}
    \dot{\phi}^2_{e} = V_{e}(\phi) = \frac{2k}{a_{e}^2}\,\, ,
\end{eqnarray}
which implies:
\begin{eqnarray}
    (T_{\phi})_{e} = \frac12 \dot{\phi}^2_{e} = \frac12 V_{e}(\phi)\,\, ,
\end{eqnarray}
where $T_\phi$ is the kinetic energy of the scalar field. The above relation between the potential energy and the kinetic energy of the scalar field at equilibrium is surprisingly similar to the virialization condition in Newtonian mechanics. It can be easily verified that the equilibrium conditions cannot be achieved  by the spatially flat FLRW spacetime, since for that we get trivial solutions: $\dot{\phi}_{e} = 0$ and $V_{e}(\phi) = 0$ which is meaningless. 

\subsection{Scalar field potential responsible for the end equilibrium state of the gravitational collapse}
Along with the two field equations of the scalar field (Eqs.~(\ref{rho}, \ref{p})), one can also write down the following conservation equation or Klein-Gordon for the scalar field:
\begin{eqnarray}
\dot{\rho}_\phi + 3\frac{\dot{a}}{a}\left(\rho_\phi + p_\phi\right) = 0  \implies \ddot{\phi}+ 3\frac{\dot{a}}{a} \dot{\phi} + V_{,\phi} = 0.
\end{eqnarray}
However, the above equation is not independent since it can be derived from Eqs.~(\ref{rho}, \ref{p}). Therefore, we have two equations and three unknowns: $\phi (a), V(\phi)$, and $\dot{a}(a)$. Therefore, we have the freedom to choose one free function. Here, we choose the function $\dot{a}(a)$ 
in such a way that the equilibrium conditions are satisfied. We consider following functional form of $\dot{a}(a)$:
\begin{eqnarray}
   \dot{a}(a) = \beta~ \left(f(a) - f(a_e)\right)^\alpha ~~ \forall a \in [a_e , a_0]\, , 
   \label{adota}
\end{eqnarray}
where the constant parameters $\beta < 0$, and $a_0$ is the initial value of the scale factor where $a_0 > a_e$. Here, we consider another assumption that $f(a) > f(a_e),~ \forall a > a_e$ and $f(a)$ should be $\mathcal{C}^\infty~~\forall a\in [a_e , a_0]$ . We can relax the last assumption by considering $\dot{a}(a) = \beta~ \lvert f(a) - f(a_e)\rvert^\alpha$. Since it is a collapsing scenario, we consider $\beta < 0$. The above expression of $\dot{a}(a)$ ensures  $\dot{a}\to 0$ as $a \to a_e$. Using the above expression of $\dot{a}(a)$, we get the following expression of $\ddot{a}(a)$:
\begin{eqnarray}
    \ddot{a}(a) = \alpha\beta^2 f^{\prime}(a)~ \left(f(a) - f(a_e)\right)^{2\alpha - 1} ~~ \forall a \in [a_e , a_0]\, ,
    \label{addota}
\end{eqnarray}
which implies $\alpha > \frac12$, since we need $\ddot{a}(a_e) = 0$. As it was discussed previously, the condition $\dot{a}=\ddot{a}=0$ at a finite comoving time does not imply the equilibrium state considering $a(t)$ as a smooth monotonically decreasing function of comoving time. When a collapsing system approaches the equilibrium state asymptotically, not only $\dot{a}$, $\ddot{a}$ tend to zero but all the higher order derivatives of $a$ with respect to the comoving time should show similar behavior. 
Therefore, considering the smooth behavior of $a(t,r)$ for a given value of $r$, we can define the following modified version of the general relativistic equilibrium state:
\begin{eqnarray}
  \lim_{t\to\infty} a^{(n,~0)}(t,r) = 0,~~\forall n\in \mathbb{Z}^+,
  \label{modeqcond}
\end{eqnarray}
where $a^{(n,~0)}(t,r)$ implies $n$th order partial derivative of $a(t,r)$ with respect to comoving time $t$ and zeroth order partial derivative of the same with respect to comoving radius $r$. Using the Cartan scalars, the above equation can be written as:
\begin{eqnarray}
    \lim_{t \to \infty}(D-\Delta)^n\Psi_2 = 0~ \forall r, ~~\forall n\in \mathbb{Z}^+.
\end{eqnarray}
Since here we express the time derivatives of the scale factor as a function of the scale factor (i.e., $a^{(n)}(a)$), the above definition (i.e., Eq.~(\ref{modeqcond})) of the general relativistic equilibrium state is very much crucial to differentiate the scenario where the $\dot{a}$ and $\ddot{a}$ become zero in finite comoving time from the scenario where the collapsing system asymptotically reaches an equilibrium state. 
Now, we can write down the following equilibrium condition for the above-mentioned homogeneous collapse:
\begin{eqnarray}
    a^{(n)}(a_e) = 0,~~\forall n\in \mathbb{Z}^+,
\end{eqnarray}
Using the expression of $\dot{a}(a)$ we can write:
\begin{widetext}
\begin{eqnarray}
\dot{a}&=& h(a)\,\, ,\nonumber\\
\ddot{a}&=&h(a)~\partial_a h(a)\,\, ,\nonumber\\
\dddot{a}&=& h^2(a)~\partial^2_a h(a) + h(a) ~(\partial_a h(a))^2\,\, , \nonumber\\
\ddddot{a}&=& h^3(a)~\partial^3_a h(a) + 4 h^2(a) \partial_a h(a)~\partial_a^2 h(a)+ h(a) ~(\partial_a h(a))^3\,\, ,\nonumber\\
\vdots\nonumber\\
a^{(n+1)}&=&\partial_t^{n}h(a)= \sum ~\frac{n!}{k_{n1}!~k_{n2}!...k_{nn}!} \partial_a^{k_n}h(a) \left(\frac{\partial_t a}{1!}\right)^{k_{n1}}\left(\frac{\partial^2_t a}{2!}\right)^{k_{n2}}...\left(\frac{\partial^m_t a}{m!}\right)^{k_{nm}}...\left(\frac{\partial^n_t a}{n!}\right)^{k_{nn}}\,
\label{delth}
\end{eqnarray}
\end{widetext}
where $k_{n1}+k_{n2}+..+k_{nm}+..+k_{nn} = k_n$,  $k_{n1}+2~k_{n2}+..+m~k_{nm}+..+n~k_{nn} = n$, and the summation is over all the partitions of $n$. To get the above expression of $a^{(n+1)}$, we use Fa\'a di Bruno's rule for the derivatives of composite functions. One can remove all the expressions of $\partial_t$ in the above expression of $a^{(n+1)}$ and can express $a^{(n+1)}$ as a function of $h(a)$ and its derivatives only, i.e., $a^{(n+1)}\equiv a^{(n+1)} (h(a), \partial_a h(a), ..., \partial_a^n h(a))$. This can be done by using the expression of $\partial_t^i a$ recursively until $i=1$ is reached. The expression of $a^{(n+1)} (h(a), \partial_a h(a), ..., \partial_a^n h(a))$ is very important since in Eq.~(\ref{adota}), $\dot{a}$ is written as a function of $a$. Now, since $h(a) = \beta~ \left(f(a) - f(a_e)\right)^\alpha ~~ \forall a \in [a_e , a_0]$, we can write:
\begin{widetext}
\begin{eqnarray}
    \partial_a h(a) &=& \beta \alpha \left(f(a)-f(a_e)\right)^{\alpha -1} f^{\prime}(a)\,\, ,\nonumber\\
    \partial_a^2 h(a) &=&\beta \alpha (\alpha -1) \left(f(a)-f(a_e)\right)^{\alpha -2} (f^{\prime}(a))^2 + \beta \alpha  \left(f(a)-f(a_e)\right)^{\alpha -1}  f^{\prime\prime}(a)\,\, , \nonumber\\
    \vdots\nonumber\\
    \partial_a^p h(a) &=& \beta\sum~ \frac{p!}{q_1!~q_2!~...q_p!}~\frac{\alpha !}{(\alpha - q) !} \left(f(a)-f(a_e)\right)^{\alpha -q} \left(\frac{\partial_a f}{1!}\right)^{q_{1}}\left(\frac{\partial^2_a f}{2!}\right)^{q_{2}}...\left(\frac{\partial^m_a f}{m!}\right)^{q_{m}}...\left(\frac{\partial^p_a f}{p!}\right)^{q_{p}}\, ,
\end{eqnarray}
\end{widetext}
where as previously done $q_1+q_2+..+q_m+..+q_p = q$,  $q_1+2~q_2+..+m~q_m+..+p~q_p = p$, and the summation is over all the partitions of $p$. Now, in order to get the expression of $a^{(i)}$ as a function of $f(a)$ and its higher order derivatives we need to put the above expression of $\partial_a^i h(a)$ in Eq.~(\ref{delth}). Now, we are ready to investigate the constrain on $\alpha$ required for an asymptotic equilibrium state. In order to get the lower-bound of $\alpha$ we need to find out the terms having the lowest power of $\left(f(a)-f(a_e)\right)$ in the expression of $a^{(n+1)}$ written in Eq.~(\ref{delth}). Followings are the lowest power of $\left(f(a)-f(a_e)\right)$ for a given value of $n$:
\begin{eqnarray}
    \dot{a}~~~ &\rightarrow&~~~\left(f(a)-f(a_e)\right)^\alpha\, \, ,\nonumber\\
    \ddot{a}~~~ &\rightarrow&~~~\left(f(a)-f(a_e)\right)^{2\alpha -1}\, \, ,\nonumber\\
    \dddot{a}~~~ &\rightarrow&~~~\left(f(a)-f(a_e)\right)^{3\alpha -2}\, \, ,\nonumber\\
    \ddddot{a}~~~ &\rightarrow&~~~\left(f(a)-f(a_e)\right)^{4\alpha -3}\, \, ,\nonumber\\
    \vdots\nonumber\\
    a^{(n+1)}~~~ &\rightarrow&~~~\left(f(a)-f(a_e)\right)^{(n+1)\alpha -n}\, \, ,
\end{eqnarray}
where one can get the lowest power of $\left(f(a)-f(a_e)\right)$ for $a^{(n+1)}$ by using the expression of $\partial_a^p h(a)$ in the above expression of $\partial_t^n h(a)$. Therefore, at equilibrium state (i.e., $a = a_e$)
\begin{eqnarray}
  \lim_{a\to a_e} a^{(n+1)} = 0 ~~\text{iff}~~   \alpha > \frac{n}{n+1}\nonumber\\ \implies
  \lim_{\substack{a\to a_e \\ n\to \infty}} a^{(n+1)} = 0 ~~\text{iff}~~   \alpha \geq 1.
\end{eqnarray} 
Therefore, $\alpha \geq 1$ ensures the null values of all the higher order derivatives of $a(t)$ in the limit $a\to a_e$, where $\dot{a}(a) = \beta~ \left(f(a) - f(a_e)\right)^\alpha ~~ \forall a \in [a_e, a_0]$. If $\alpha = \frac{n}{n+1}$ (i.e., $\alpha < 1$) then all the higher order derivatives of $a(t)$ become zero except $a^{(n+1)}$, which implies that the collapsing dynamics is not settling down a stable equilibrium state. A positive or negative value of $a^{(n+1)}(a_e)$ implies that $a^{(n)}$ is going to have positive or negative values, respectively as comoving time progresses. If $ \frac{n-1}{n}<\alpha< \frac{n}{n+1}$, all the derivatives from $\dot{a}$ to $a^{(n)}$ would become zero at $a = a_e$, however, remaining all the higher order derivatives of $a(t)$ (i.e., $a^{(n+1)}, a^{(n+2)}, \cdots a^{(\infty)}$) would blow up at $a=a_e$ which is obviously not the equilibrium scenario. Therefore, $\alpha \geq 1$ isolates equilibrium scenarios from the various other dynamics where $\dot{a}, \ddot{a}$ may become zero at a certain comoving time but do not imply an equilibrium state. Therefore, from now on, we are going to consider only  $\alpha \geq 1$.

Using the expression of $\dot{a}$ and $\ddot{a}$ in Eq.~(\ref{adota}) and Eq.~(\ref{addota}), respectively we get:
\begin{widetext}
\begin{eqnarray}
    (\partial_a\phi)^2 &=& -\frac{2\alpha~f^{\prime}(a)}{a\left(f(a) - f(a_e)\right)}+\frac{2}{\beta^2~a^2\left(f(a) - f(a_e)\right)^{2\alpha}}+\frac{2}{a^2}\,\, , ~~\forall a\in [a_e , a_0] ,\\
    V(a) &=& \frac{\alpha\beta^2f^{\prime}(a)}{a}\left(f(a) - f(a_e)\right)^{2\alpha -1}+\frac{2\beta^2}{a^2}\left(f(a) - f(a_e)\right)^{2\alpha}+\frac{2}{a^2}\,\, ,~~\forall a\in [a_e , a_0] .
    \label{Va1}
\end{eqnarray}
\end{widetext}
The above equations show the functional dependence of $\partial_a\phi$ and the scalar field potential $V$ on $a$ for the collapsing dynamics where collapsing system asymptotically reaches the equilibrium state.
Since $\alpha\geq 1$, asymptotically $V(a_e) = \frac{2}{a_e^2}$ and $\partial_a\phi \to\pm\infty$. Close to the equilibrium state i.e., when $a\to a_e$ we can write:
\begin{eqnarray}
\label{delaphi}
    \partial_a\phi &\approx& \frac{\pm\sqrt2}{\beta~a\left(f(a) - f(a_e)\right)^{\alpha}}\,\, ,\\
    V(a) &\approx& \frac{2}{a^2}\, .
    \label{Va2}
\end{eqnarray}
Using the above approximate solutions of $\partial_a\phi$ and $V(a)$, we can analytically show the expression of scalar field potential as a function of $\phi$ i.e., $V(\phi)$ which would be applicable near equilibrium state only. From the above approximate solution of $\partial_a\phi$, we can write:
\begin{eqnarray}
\phi(a)\approx\pm\frac{\sqrt2}{\beta}\int_a^{a_0}\frac{da}{a\left(f(a) - f(a_e)\right)^{\alpha}}.
\end{eqnarray}
For analytical expression of $\phi(a)$, lets consider $f(a) = a$. Now, for this simple form of $f(a)$ we get the following solutions of $V(\phi)$ near the equilibrium states:
\begin{eqnarray}
\text{when}~~\alpha = 1 :~~    V(\phi) &\approx& \frac{2}{a_e^2}\left(1-2\eta ~e^{\frac{\beta a_e}{\sqrt2}\phi}\right) ~~\textbf{for}~~ \phi > 0 \,\, ,\nonumber\\
    &\approx& \frac{2}{a_e^2}\left(1-2\eta~ e^{-\frac{\beta a_e}{\sqrt2}\phi}\right) ~~\textbf{for}~~ \phi < 0 \,\, ,\label{V1}\nonumber\\
    \\
    \text{when}~~\alpha > 1 :~~    V(\phi) &\approx& \frac{2}{a_e^2}\left(1-\frac{2~\zeta}{\phi^{\frac{1}{\alpha -1}}} \right) ~~\textbf{for}~~ \phi > 0 \,\, ,\nonumber\\
    &\approx& \frac{2}{a_e^2}\left(1-\frac{2~\zeta}{(-\phi)^{\frac{1}{\alpha -1}}}\right) ~~\textbf{for}~~ \phi < 0 \,\, ,
    \label{V2}
    \nonumber\\
\end{eqnarray}
where $\eta$ is the integration constant, $\zeta =\left[ -\frac{\beta\left(\alpha-1\right)}{\sqrt2 a_e^\alpha}\right]^{-\frac{1}{\alpha -1}}$, and since $\beta<0, \text{and}~ \alpha>1$, $\zeta > 0$. It can be verified that in the limit $a\to a_e$, as $\phi\to \pm\infty$, $V(\phi)$ approaches to $\frac{2}{a_e^2}$ asymptotically. It should be noted that the above approximate expressions of $V(\phi)$ are only valid close to the equilibrium state, however, the expression of $V(a)$ given in Eq.~(\ref{Va1}) is valid throughout the gravitational collapse, i.e., for $a\in [a_e, a_0]$.
\begin{figure*}
\centering
\includegraphics[scale=0.6]{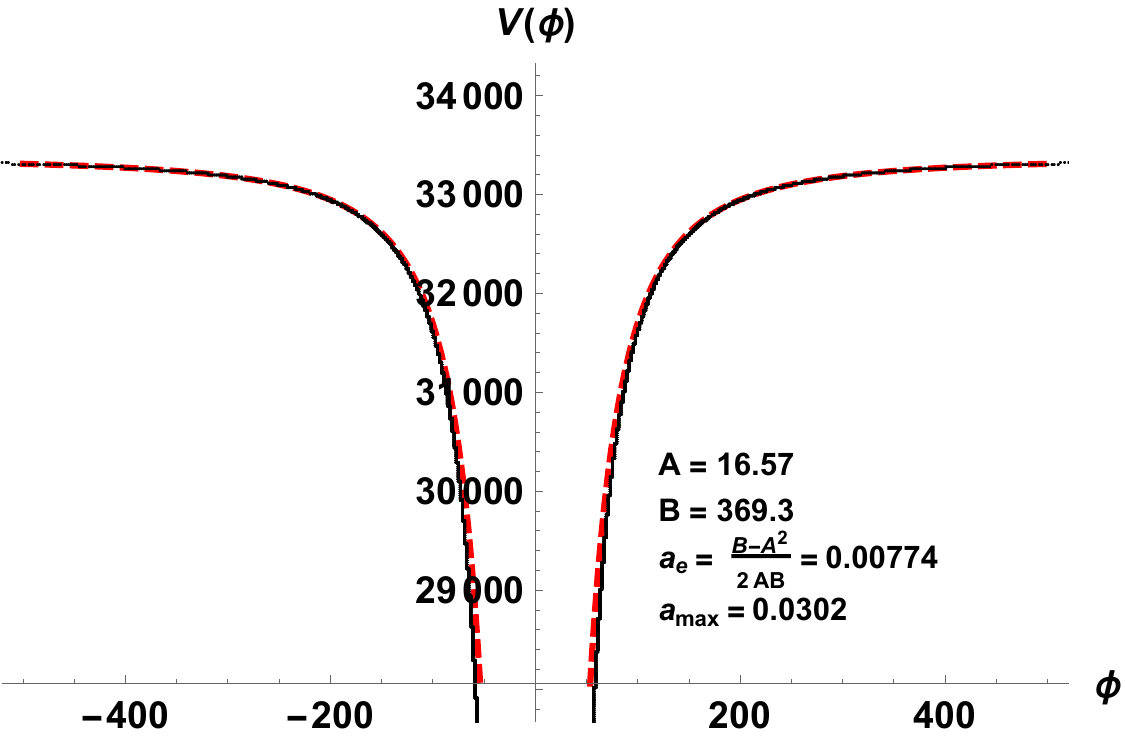}
\caption{Figure depicts the scalar field potential near the equilibrium where the dotted red curve corresponds to the approximate potential near the equilibrium and the black solid line is for the exact solution of potential.}
\label{potentialscalar}
\end{figure*}
\section{Scalar Field Potential in a Bonafide Model of Gravitational Collapse Leading to an Equilibrium State}
\label{sec4}
In this section, we derive the scalar field potential that is necessary for achieving an equilibrium state at the end of the gravitational collapse of homogeneous matter in a bonafide model. The model we refer to is extensively discussed in \cite{Dey:2023qdt}. Here, we employ the same dynamical model and obtain the scalar field potential that is required, assuming that the collapsing fluid is composed of the minimally coupled scalar field. In \cite{Dey:2023qdt}, the gravitational collapse of homogeneous matter is modeled by considering a closed Friedmann-Lemaître-Robertson-Walker (FLRW) spacetime described by the following metric:
\begin{eqnarray} \label{FLRW1}
ds^{2}=-dt^{2}+\frac{a^{2}(t)}{1-r^{2}}dr^{2}+r^{2}a^{2}(t)~d\Omega^2\,\, .\nonumber\\ 
\end{eqnarray} 
To capture a collapsing scenario leading to an equilibrium state, the collapse model incorporates the following expression for $\dot{a}$:
\begin{eqnarray}
 \dot a(t_r)=-{A(t_r-1)\over(1+B(t_r-1)^2)^2}~,~t_r = {t\over t_{\rm max}}~,
\label{fdotexp}
\end{eqnarray}
Here, $A$ and $B$ represent real positive constants, while $t_{\rm max}$ denotes the time when the spherical homogeneous region begins to collapse under its own gravitational force. At $t = t_{\rm max}$, the function $a(t)$ reaches its maximum value, denoted as $a_{\rm max}$.
In \cite{Dey:2023qdt}, the bonafide model is employed to describe a cosmological scenario where spherically symmetric over-dense regions of dark matter and dark energy can achieve an equilibrium configuration at a specific cosmological length and time scale. In this dynamics, the over-dense region initially expands with the background, albeit at a slower rate. However, at a certain time $t = t_{\rm max}$, the expansion comes to a halt, and the over-dense region starts collapsing. Consequently, at $t = t_{\rm max}$, the value of $\dot{a}$ becomes zero which can be verified from Eq.~(\ref{fdotexp}). Using the above expression of $\dot{a}$, we can get the following expression of the scale factor $a(t_r)$:
\begin{eqnarray}
 a(t_r)=a_e+\frac{A}{2 B \left(1+B (t_r-1)^2\right)}\, ,
\label{fex}
\end{eqnarray}
where $a_e$ is the value of the scale factor at equilibrium which has the following expression considering zero pressure at time $t = t_{\rm max}$:
\begin{eqnarray}
a_e={B-A^2\over 2AB}\,.
\label{fzero}
\end{eqnarray}
where we need $B>A^2$ to get a positive value of $a_e$. From Eqs.~(\ref{fdotexp}, \ref{fex}), one can see that the equilibrium configuration can be achieved by the collapsing system at asymptotic time. Using Eqs.~(\ref{fdotexp}, \ref{fex}), we can write down the $\dot{a}$ as a function of $a$:
\begin{equation}\label{scaledey}
    \Dot{a}=-\frac{4B^2}{A} {\left[\frac{A}{2B^2}\frac{1}{(a(t)-a(t_e))}-\frac{1}{B}\right]}^{\frac{1}{2}}{(a(t)-a(t_e))}^2\,\, ,
\end{equation} 
which in the limit $a\to a_e$ becomes:
\begin{eqnarray}
\dot{a}(a)\approx\beta \left(a~-~a_e\right)^{\frac32}\,\, ,
\end{eqnarray}
where $\beta = -\frac{2\sqrt2}{\sqrt{A}}B$. As discussed in the previous section, in this case, $\alpha = \frac32$ which is greater than one, and therefore, this collapsing system has an end equilibrium state. It can be verified for this present scenario, $\lim_{t\to\infty} a^{(n,~0)}(t,r) = 0,~~\forall n\in \mathbb{Z}^+$ which implies the existence of end equilibrium state. Using Eqs.~(\ref{delaphi}, \ref{Va2}), and the expression of $V(\phi)$ near the equilibrium in Eq.~(\ref{V2}), we get:
\begin{eqnarray}
\phi(a) &\approx & \pm\frac{2\sqrt2}{\beta a_e}\frac{1}{\left(a~-~a_e\right)^{\frac12}}\,\, ,\\
V(\phi)&\approx &\frac{2}{a_e^2}\left(1-\frac{2\zeta}{\phi^{2}}\right)\,\, ,
\end{eqnarray}
where $\zeta = \frac{8B}{A^2}\frac{1}{\left(\frac{B}{A^2}-1\right)^3}$. Since $B> A^2$, $\zeta > 0$. In Fig.~(\ref{potentialscalar}), we can see that the approximate solution of the potential of the scalar field (depicted by dotted red curve) almost coincides the exact solution (depicted by black solid line) near the equilibrium configuration. Here, we consider $A = 16.57, B= 369.3$ which are relevant to the cosmological scenario discussed in \cite{Dey:2023qdt}. Therefore, the system starts collapsing at $a_{\rm max} = 0.0302$ and asymptotically reaches the equilibrium state where $a_e = 0.00774$.

\section{Conclusion}
In this paper, we study the gravitational collapse of a minimally coupled scalar field with non-zero potential, and we obtain a class of scalar field potentials that lead to an equilibrium end state of gravitational collapse.  
In order to do that we first express the equilibrium condition in the following way
\begin{eqnarray}
    a^{(n)}(a_e) = 0,~~\forall n\in \mathbb{Z}^+,
\end{eqnarray}
which is just a different form of the condition stated in Eq.~(\ref{eqlb}). While obtaining the potential of the scalar field,
we express every dynamical variable as a function of the scale factor. Now at equilibrium, the scale factor becomes $a_e$. A scale factor can reach a specific value at a given comoving time. However, it becomes challenging to ascertain whether that particular value of the scale factor corresponds to the equilibrium state, as all dynamical variables are expressed in terms of the scale factor rather than the comoving time. To address this, one can verify the higher-order derivatives of the scale factor. The null values of all higher-order derivatives along with the first and second at a certain scale factor value $a_e$ ensure the existence of an end equilibrium state of the collapsing system. The null values of the first and second-order derivatives of the scale factor at an asymptotic comoving time do imply the existence of an equilibrium state. However, when one expresses all the dynamical variables in terms of scale factor, null values of the first two derivatives at a certain value of scale factor $a_e$ do not imply an equilibrium state, since there is no information of comoving time (e.g. $a= a_e + (t_e - t)^n,~ \forall n >2$, the first two derivatives become zero at $a = a_e$ but the collapsing system has no equilibrium state at $a = a_e$). We have to show null values of all the higher-order derivatives at a certain value of $a$.  We show the lower limit of a parameter $\alpha$ for which all the higher order derivatives of the scale factor become zero at $a = a_e$. 
One can use the comoving time to express an equilibrium dynamic. However, with that procedure, one cannot cover the large class of collapsing dynamics that can reach the equilibrium state. In \cite{Dey:2023qdt}, the authors show one example of such a dynamics, where the scale factor has a particular functional dependence on comoving time. By expressing all variables as functions of the scale factor $a$, we eliminate the need to determine the dependence of $a$ on the comoving time, except for one crucial constraint: all higher-order derivatives of $a$ must become zero at $a = a_e$. In this way, we cover a large class of collapsing dynamics leading to an end equilibrium state.
Additionally, if one expresses everything in terms of comoving time then it may be very difficult to get an analytical form of $V(\phi)$ and then we would have to show the potential by a numerical plot which is not of our interest in this paper.

In Sec.~(\ref{sec4}), we derive the scalar field potential required to achieve an equilibrium state at the conclusion of the gravitational collapse of homogeneous matter in a bonafide model discussed in \cite{Dey:2023qdt}. We show that the scalar field potential corresponding to the example belongs to the class of potentials discussed in Sec.~(\ref{sec3}). As stated before, in this paper, we refrain from presenting the scalar field as a candidate for dark matter or dark energy. Instead, our primary objective lies in deriving the potential of the scalar field, where it plays a crucial role as the constituent of the collapsing matter. At a certain cosmological length and time scale, if the gravitational collapse of a certain homogeneous matter distribution consisted of a scalar field approach towards the general relativistic equilibrium state, then after a large comoving time, the potential of the scalar field may have a similar functional form given in Eqs.~(\ref{V1}, \ref{V2}). Now, the next step would be to find out some cosmological implications of the aforementioned scalar field potential. However, we have deferred this aspect to future work, acknowledging the need for further exploration and analysis in this direction.
\label{sec5}

\section{Acknowledgement}
DD would like to acknowledge the support of
the Atlantic Association for Research in the Mathematical Sciences (AARMS) for funding the work.

\end{document}